\documentclass[newstyle,twocolumn,proceedings]{rmaa}

\usepackage{rmaacite}

\newcommand\etal{et al.}
\newcommand\lya{Ly$\alpha$}

\newcommand\civ{\hbox{C~\small$\rm IV$}}
\newcommand\heii{\hbox{He~\small$\rm II$}}
\newcommand\nv{\hbox{N~\small$\rm V $}}
\newcommand\siiv{\hbox{Si~\small$\rm IV$}}
\newcommand\oiii{\hbox{O~\small$\rm III]$}}

\title{Deep VLT spectroscopy of High Redshift Radio Galaxy MRC~2104--242: 
Evidence for a metallicity gradient}
\author{R. Overzier, H. R\"ottgering, J. Kurk and C. De Breuck
  \affil{Sterrewacht Leiden, The Netherlands} }

\fulladdresses{
\item R. Overzier, H. R\"ottgering, J. Kurk and C. De Breuck: Sterrewacht Leiden, P.O. Box 9513, 2300 RA, Leiden, The Netherlands (overzier@strw.leidenuniv.nl).}

\shortauthor{Overzier, R\"ottgering, Kurk and De Breuck}
\shorttitle{Radio Galaxy MRC 2104--242}


\abstract{In this contribution we will present deep VLT spectroscopy 
observations of the giant emission line halo around the $z=2.49$ radio galaxy 
MRC 2104--242. The morphology of the halo is dominated by two spatially 
resolved regions. Ly$\alpha$ is extended by $>$12$\arcsec$\ along the radio 
axis, C IV $\lambda\lambda$\ 1549 and He II $\lambda$\ 1640 are extended by 
$\sim$8$\arcsec$. The overall spectrum is typical for that of high redshift 
radio galaxies. Interestingly, N V $\lambda$\ 1240 
is present in the spectrum of the region associated with the center of the 
galaxy hosting the radio source, the northern region, while absent in the 
southern region. Using a simple photoionization model, the difference in 
N V $\lambda$ 1240 emission can be explained due to a metallicity gradient 
within the halo. This is consistent with a scenario in which the halo 
is formed by a massive cooling flow or originates from the debris of 
the merging of two or more galaxies. However, also other mechanisms such as  
jet-cloud interactions or starburst-winds could be important.}

\listofauthors{R.~Overzier, H.~R\"ottgering, J.~Kurk \& C.~De~Breuck}
\indexauthor{Overzier, R.}
\indexauthor{R\"ottgering, H.}
\indexauthor{Kurk, J.}
\indexauthor{De Breuck, C.}

\begin{document}

\maketitle


\section{Introduction}
\label{sec:introduction}

High redshift radio galaxies (HzRGs) are often surrounded by giant 
halos of ionized gas, which radiate luminous emission lines in the rest frame 
UV/optical part of the spectrum (see \pcite{mc93}\ for a review). 
The typical emission line spectrum of 
HzRGs can best be explained assuming photoionization by a hidden quasar, 
with in some cases an additional contribution due to shockionization by 
jet-cloud interactions \cite{best00}. 

Because HzRGs are believed to be the progenitors of massive elliptical 
galaxies (e.g. \pcite{best98}) and have many observable components, 
they are a good tool for studying the formation of massive galaxies. 
An important question related to this concerns the origin of 
the halo. If the halo gas originates from outside the radio galaxy, it 
could be associated with the debris from galaxies merging \cite{heck86} 
or it could originate in a massive cooling flow from a cluster type 
halo \cite{crawfab96}. Alternatively, the gas could have been driven out 
by a starburst-wind or by shocks associated with the radio source. 
Studying the properties of the gas in detail may help to make a distinction 
between these or other scenarios.
  
In this contribution, we will first give a review of the object of our study, 
radio galaxy 2104--242 at $z=2.49$, in \S\ 2. 
Then, in \S\ \ref{sec:spectroscopy}, 
we present the results of spectroscopic observations with the VLT and we 
will show that we have found evidence for a metallicity gradient in the 
extended emission line region in 
\S\ \ref{sec:metallicity}. To conclude, we will discuss the origin of the 
halo of 2104--242 in \S\ \ref{sec:discussion}. 


\section{MRC 2104--242}
\label{sec:2104}

Radio source 2104--242 is identified with a galaxy at $z=2.49$ 
and is one of the brightest known HzRGs in \lya\ \cite{mc90}.  
Narrowband \lya\ images show a total extent of\ $>$ 12\arcsec\ 
(i.e. 136 kpc\footnote[1]{We adopt H$_{0}=50$ 
km s$^{-1}$ Mpc$^{-1}$, q$_{0}=0.1$. At $z=2.49$ this implies a 
linear size scale of 11.3 kpc arcsec$^{-1}$.}) 
distributed in two distinct regions separated by\ 
$\sim$6\arcsec. Because of the brightness and the spectacular morphology 
of the halo at the considerable redshift of $z=2.49$, 2104--242 has been 
the subject of a number of different observational programs including 
HST-imaging \cite{pen99}\ and VLT spectroscopy 
\cite{over01}. 

\begin{figure}
\includegraphics[width=\columnwidth]{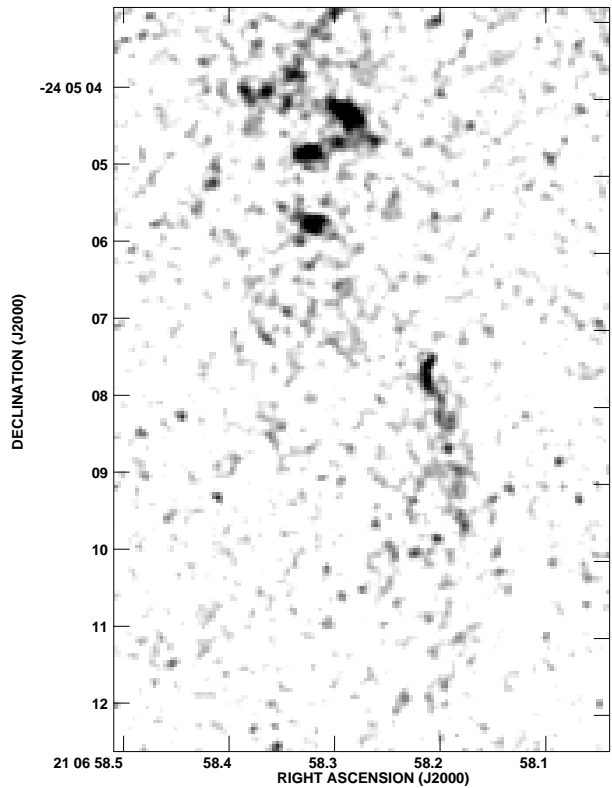}
\caption[]{WFPC2 V-band image of 2104--242 in grey scale \cite{pen99}. 
Some of the components in the top of this image are confused with the 
residuals from a spike of a nearby star.}
\end{figure}

HST--WFPC2 V--band images of 2104--242 show that this  galaxy actually 
consists of several smaller components, presumably in the process of 
galaxy-merging (see Fig. 1). One of the bright components hosted by the 
northern \lya\ region is prominent in near-infrared emission and therefore 
it is assumed to be the center of the galaxy hosting the radio source 
\cite{pen99}. 
The continuum of the nucleus and the other components in 
this region are not spatially resolved in our 
spectra, so we will refer to the whole region as the northern region. 
The southern \lya\ region is associated with the narrow filamentary component 
of $\sim$2\arcsec\, clearly seen in Fig. 1. This component is oriented in 
close alignment with the direction of the radio axis (see Fig. 2 for an 
HST--NICMOS image overlayed with contours of the narrowband \lya\ halo). 

Spectroscopy shows that both \lya\ regions have large FWHM 
($\sim$$1000-1500$\ km s$^{-1}$), large rest-frame
equivalent widths (330 and 560 \AA) and a velocity difference 
of $\sim$500 km s$^{-1}$ \cite{mc96,koek96,vil99}. 
The two regions also emit other lines and faint continuum. 

\begin{figure}[t]
\includegraphics[width=\columnwidth]{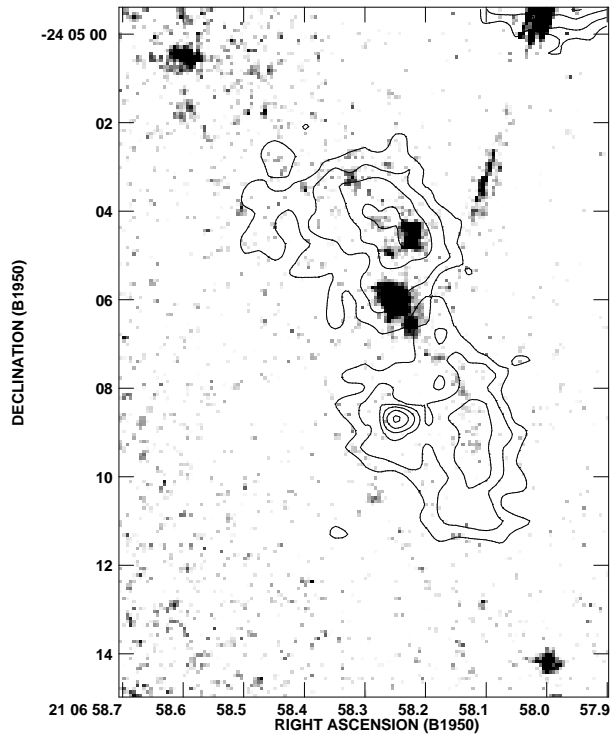}
\caption[]{NICMOS image of 2104--242 in grey scale overlayed with 
contours of the narrow band Ly$\alpha$\ emission  \cite{pen99}. 
The narrow feature in the upperright of this image is a spike of a 
nearby star.}
\end{figure}


\section{VLT spectroscopy} 
\label{sec:spectroscopy}

We used the FORS1 spectrograph on the 8.2m VLT Antu 
telescope (ESO-Chile) with a 1\arcsec\ wide slit.
The slit was positioned along the bright components and 
the filamentary structure seen in Figs. 1 and 2. 
The exposure time was 3$\times$3600s. 

Fig. 3 shows the two-dimensional spectra of \lya, \civ\ 
and \heii.  \civ\ and \heii\ are extended by $\sim$8$\arcsec$ along the 
radio axis and Ly$\alpha$ is extended by $>$12$\arcsec$. We have compared 
the kinematic structure of these 

\begin{figure}
\begin{minipage}{\columnwidth}
\includegraphics[width=6cm,height=6cm]{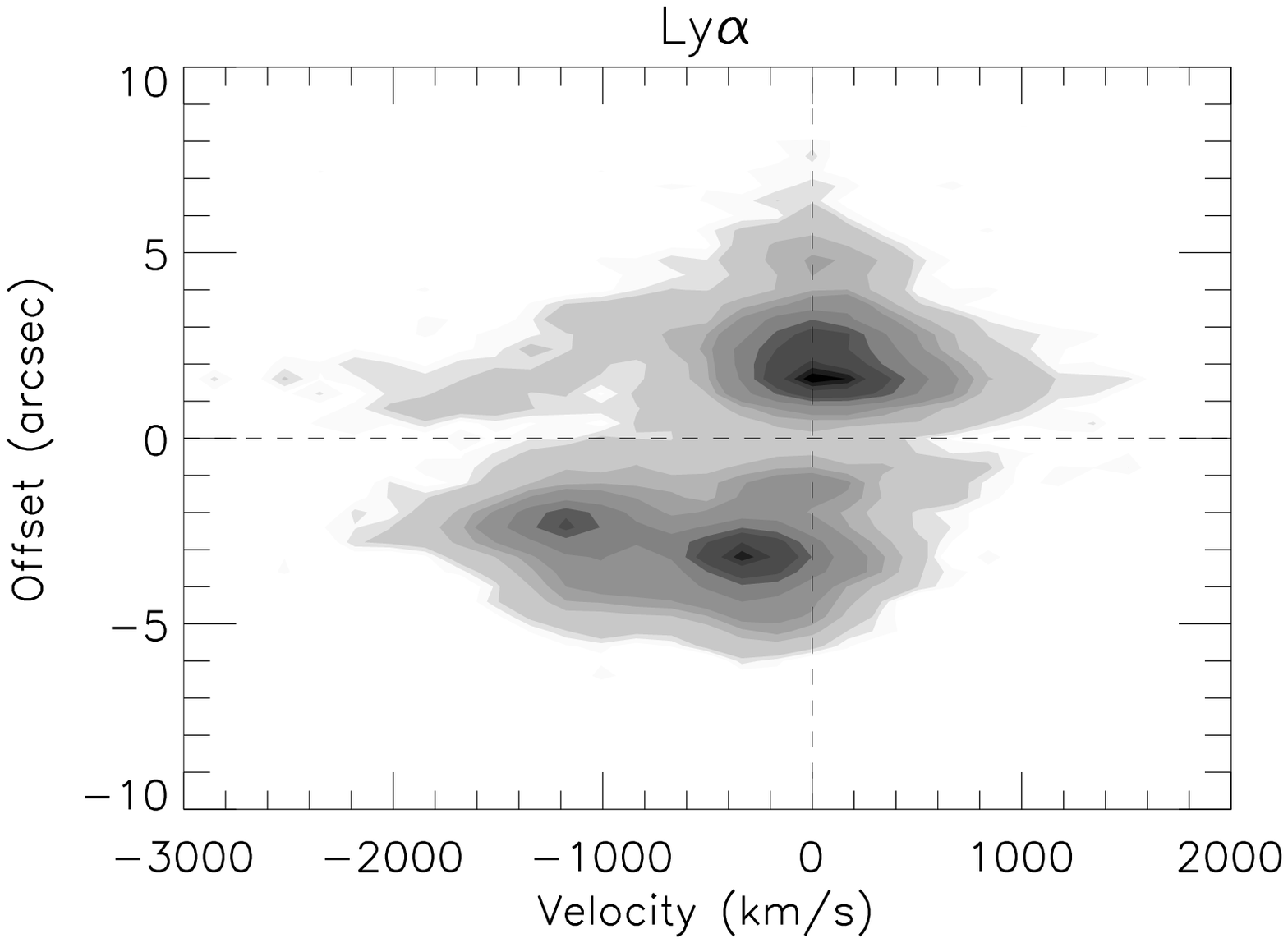}
\includegraphics[width=6cm,height=6cm]{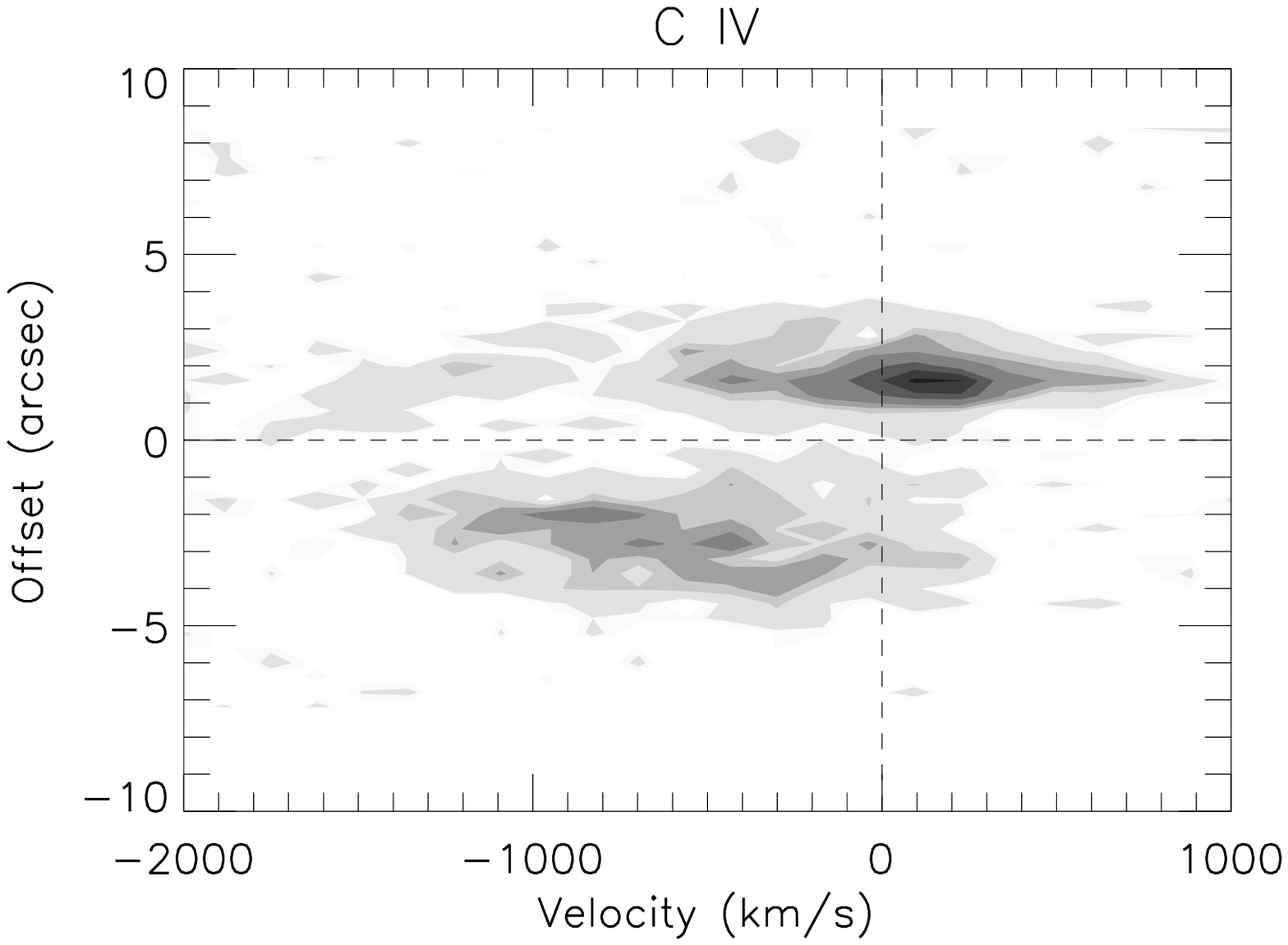}
\includegraphics[width=6cm,height=6cm]{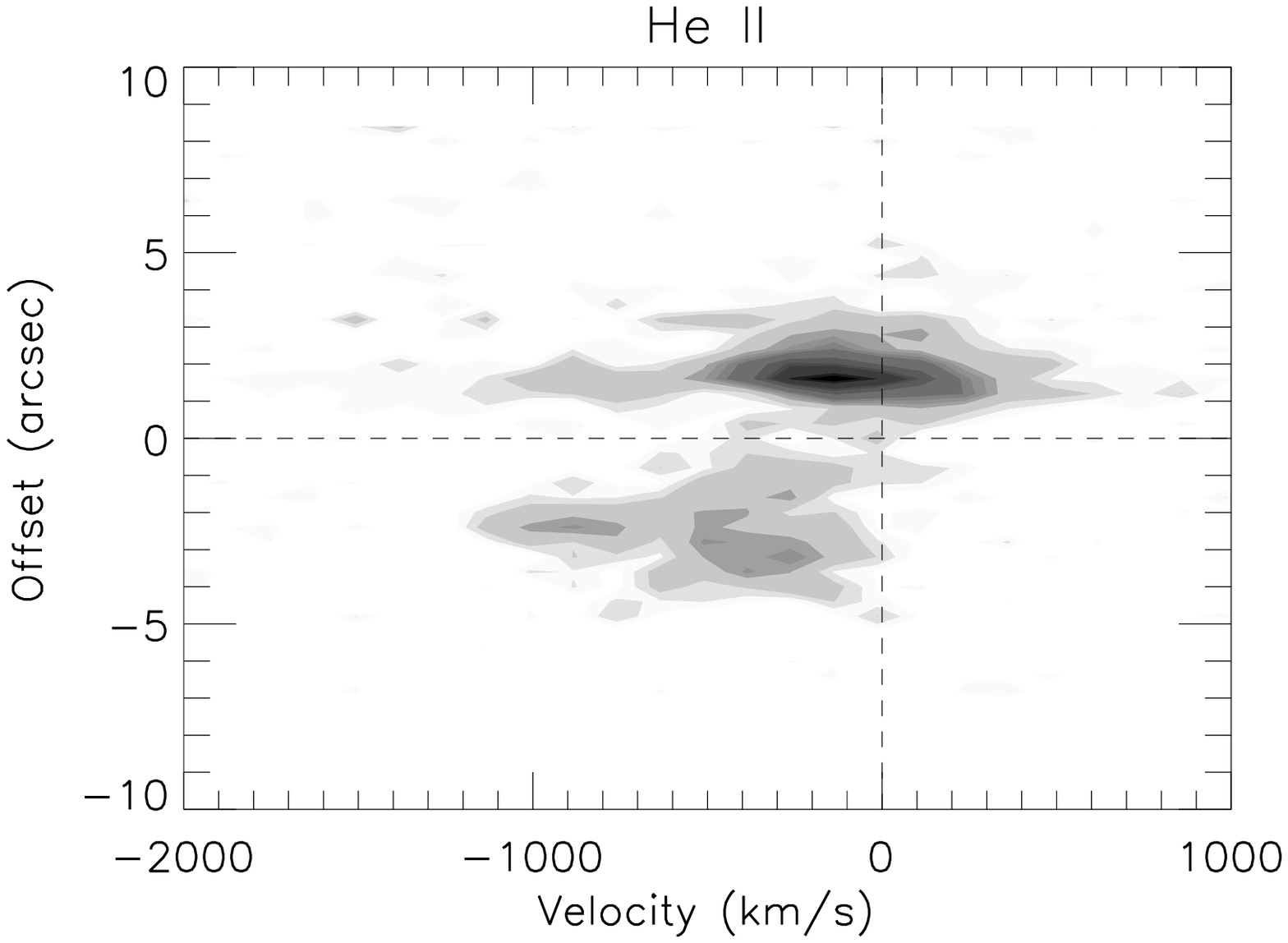}
\caption{The two-dimensional emission line structures of \lya, \civ\ 
and \heii. Offset zero was chosen in between the northern and southern 
regions. Velocity zero corresponds to the peak \lya\ emission in the 
northern region.}
\end{minipage}
\end{figure}
\begin{figure}
\begin{minipage}{\columnwidth}
\includegraphics[width=\columnwidth]{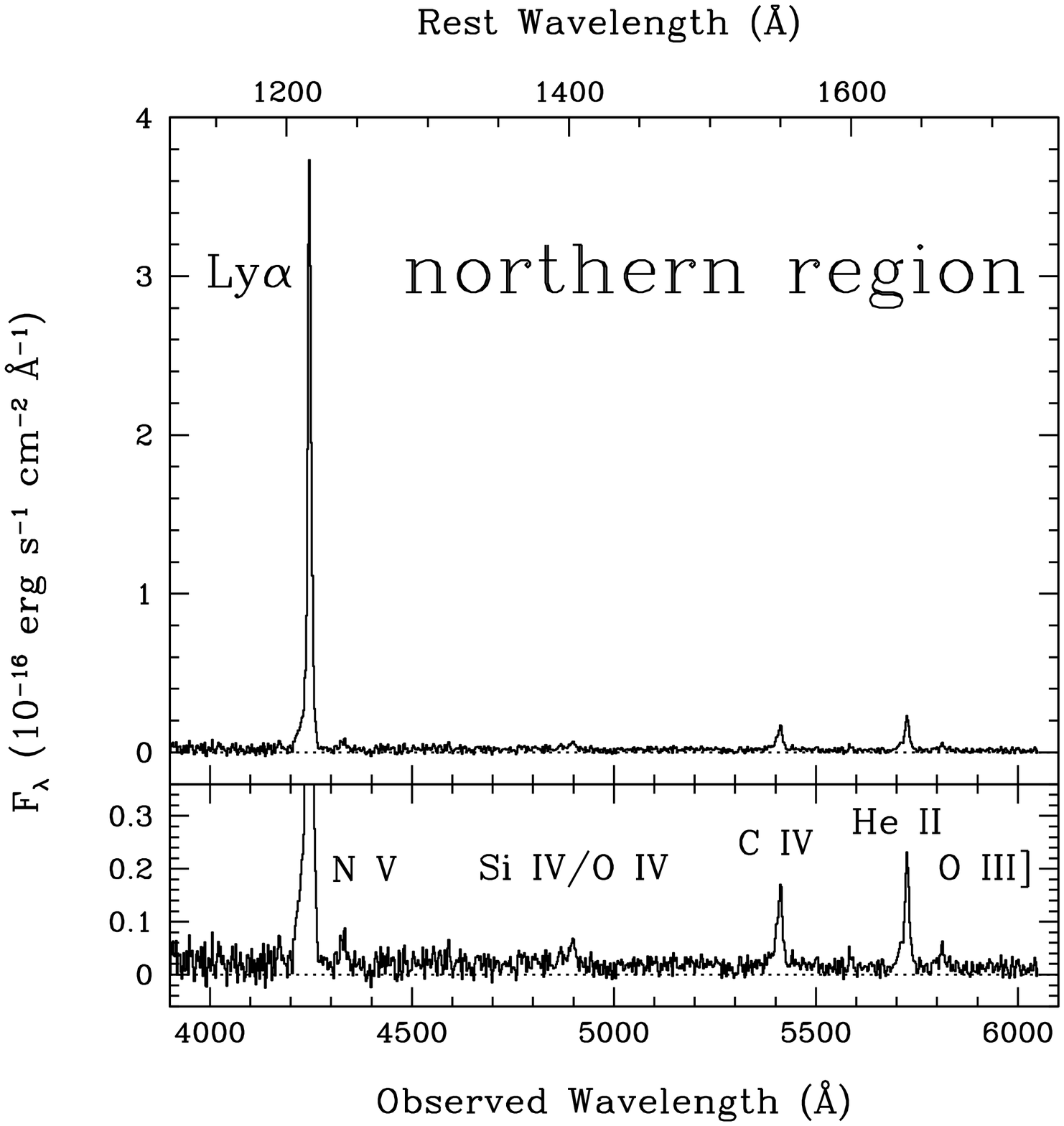}
\caption[]{Spectrum of the northern region of emission of 2104--242.}
\end{minipage}
\begin{minipage}{\columnwidth}
\includegraphics[width=\columnwidth]{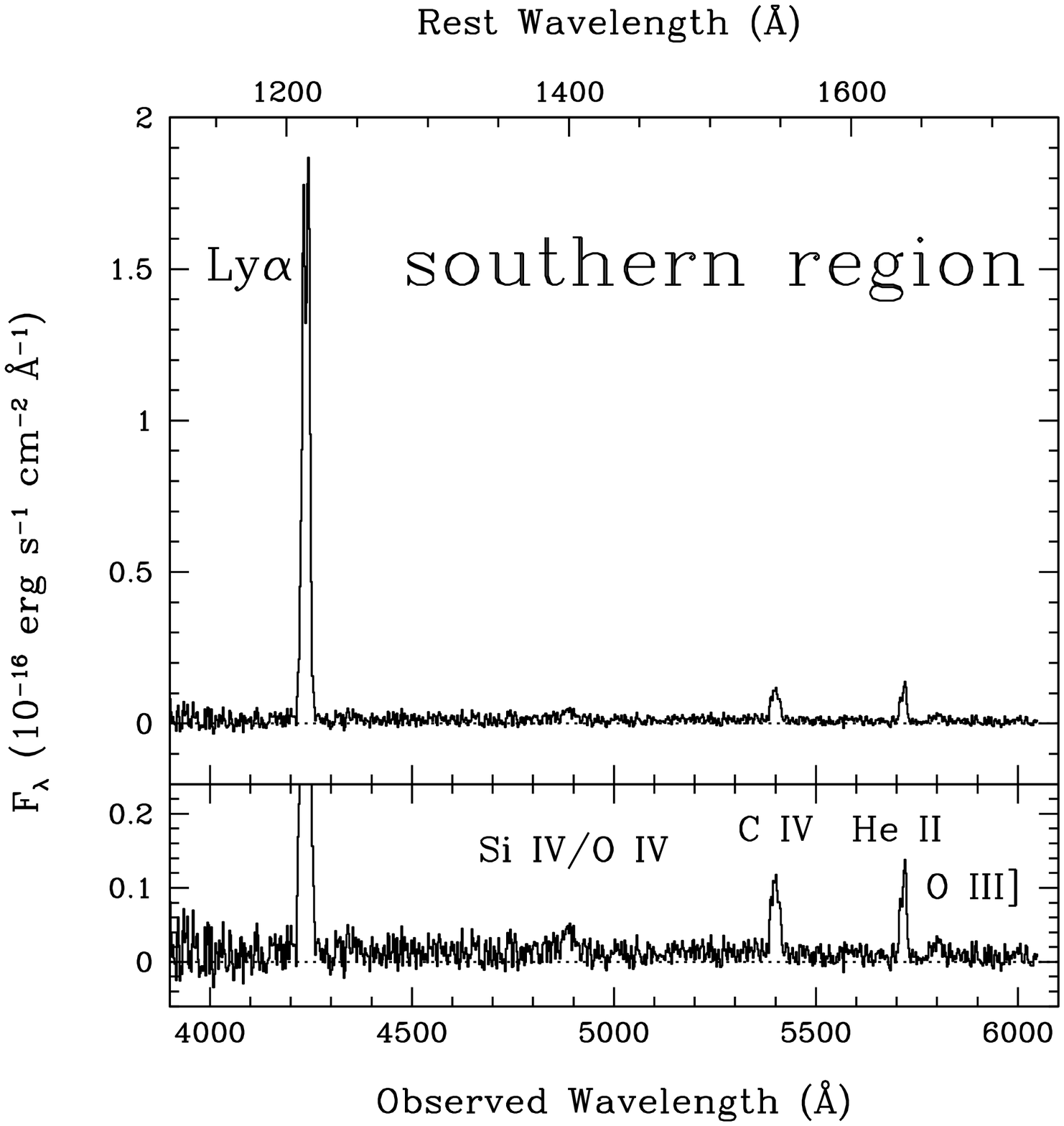}
\caption[]{Spectrum of the southern region of emission of 2104--242.}
\end{minipage}
\end{figure}

\noindent
lines. The peak of the \civ\ emission in the northern 
region is redshifted with respect to that of \lya\ by $\sim$100 km s$^{-1}$, 
while that of \heii\ is blueshifted by $\sim$150 km s$^{-1}$.
In the southern region, \lya\ shows two separate peaks shifted blueward from 
the northern region by $\sim$1000 and $\sim$500 km s$^{-1}$. This two-peak 
distribution is also seen in \civ\ and \heii. The fact that it is observed 
in \heii\ could indicate kinematical substructure in the halo, because \heii\ 
is unsusceptible to absorption. Therefore, we conclude that the dip in 
the \lya\ profile is not due to absorption. 

Figs. 4 and 5 show the one-dimensional spectra of the northern 
and southern region of the emission line halo. Both regions show 
bright \lya, \civ\ and \heii\ and weak \siiv\ and \oiii. Interestingly, 
\nv\ is detected in the northern region, but absent in the southern region. 
Within the errors, the emission line ratios of the two regions are the same, 
only those involving \nv\ are discrepant. The \nv/\civ\ and \nv/\heii\ line 
ratios are at least 4 and 3 times higher in the northern region compared to 
the southern. In \S\ 4 we will show that this is evidence for a metallicity gradient within the halo.


\section{Evidence for a metallicity gradient?}
\label{sec:metallicity}

\scite{ver00} found that HzRGs follow a sequence in \nv/\civ\ vs. 
\nv/\heii, parallel to the relation defined by the broad line 
regions (BLRs) of quasars found by \scite{hamfer93}. 
\scite{hamfer93} showed that this sequence can be explained by a different 
metallicity of the BLRs. This could be due to variation in the evolutionary stage 
and intensity of the associated starburst. \scite{ver00} found that the 
\nv\ correlation and the strong \nv\ emission 
in some HzRGs (e.g. \pcite{ojik94}) are best explained by a model of 
photoionization and variation of metallicity. Therefore, we conclude that 
the difference in \nv\ emission from the two regions can be explained only 
by a metallicity gradient within the halo. Using a metallicity sequence 
with quadratic nitrogen enhancement (N\ $\propto$\ Z$^{2}$) from 
\scite{ver00} we find a metallicity of Z\ $\approx$\ 1.5 Z$_{\odot}$ for the 
northern region and an upper limit of Z\ $\leq$\ 0.4 Z$_{\odot}$\ for the 
southern (Fig. 6).

\section{Discussion}
\label{sec:discussion}

The supersolar metallicity of the gas associated with the central 
part of the galaxy implies that 2104--242 has experienced a period of 
intense star formation. Assuming that the difference in \nv\ emission 
found within the halo is indeed due to a metallicity gradient, 
the emitting gas near the center and the gas further out are in 
different stages of evolution.

It has been suggested that the infall of gas by massive cooling flows 
is an important process in galaxy formation \cite{crawfab96}. 
In this scenario, gas cools from a primordial 
halo surrounding the radio source and provides the material from which the 
galaxy is made. \scite{fardal00} examined cooling radiation from forming 
galaxies, estimating \lya\ line luminosities of high redshift systems. 
They find that a significant amount of the extended \lya\ emission can indeed 
arise from cooling radiation. 

\begin{figure}
\includegraphics[width=\columnwidth]{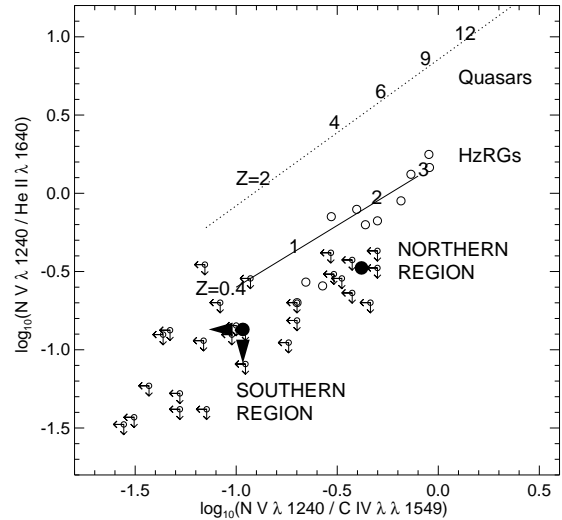}
\caption[]{\nv/\heii\ vs. \nv/\civ. The dotted line represents the 
metallicity sequence defined by quasars (\pcite{hamfer93}), 
with the numbers indicating the metallicity in solar units. 
The solid line represents a metallicity sequence with N\ $\propto$\ Z$^{2}$ 
(ionization parameter U$=$0.035, power law spectral index $\alpha$$=-$1.0) 
from \scite{ver00}. The two regions of 2104--242 are 
indicated. Small open circles indicate radio galaxies from the sample of 
\scite{db00}.}
\end{figure}

However, because the two regions of 2104--242 both show 
emission lines from other elements than H and He, at least some of 
the gas must already have been processed in stars in the past. 
Therefore, other mechanisms may be needed to explain the origin of the halo.
The HST images of 2104--242 \cite{pen99} support the idea that galaxies are formed by a process 
of hierarchical buildup \cite{kauf95}. We have also found evidence for 
kinematical substructure within the halo. Therefore, the emission line halo 
may be the result of gas associated with intense merging of galaxies. 

Alternatively, the gas in the halo could be the result of 
jet-cloud interactions, inducing extreme, non-gravitational motions or it 
could have been expelled by a superwind following an 
enormous starburst. \scite{binette00} showed that 
radio galaxy 0943--242 is surrounded by a gas shell of low 
metallicity. They conclude that this gas has been expelled from 
the parent galaxy during the initial starburst at the onset of its formation.
 
We conclude that the emission line ratios are well explained by a 
combination of photoionization and a metallicity gradient. This is 
consistent with scenarios in which the halo is formed by gas falling 
onto the radio galaxy located at the center of a forming cluster or 
by gas associated with intense galaxy merging, but other mechanisms such as 
jet-cloud interactions or starburst-winds cannot be ruled out. 

\acknowledgements 
We are very grateful to Wil van Breugel and Laura Pentericci 
for productive discussions.


\end{document}